# Fabrication and Optical Properties of a Fully -Hybrid Epitaxial ZnO-Based Microcavity in the Strong-Coupling Regime

L. Orosz[1,2], F. Réveret[1,2,*], S. Bouchoule[3], J. Zúñiga-Pérez[4], F. Médard[1,2], J. Leymarie[1,2], P. Disseix[1,2], M. Mihailovic[1,2], E. Frayssinet[4], F. Semond[4], M. Leroux[4], M. Mexis[5,6], C. Brimont[5,6], and T. Guillet[5,6]

[1] *LASMEA, Clermont Université, Université Blaise Pascal, BP 10448, F-63000 Clermont-Ferrand, France*
[2] *CNRS, UMR 6602, 24 Avenue des Landais, F-63177 Aubière Cedex, France*
[3] *LPN–CNRS, Route de Nozay, F-91460 Marcoussis, France*
[4] *CRHEA–CNRS, Rue Bernard Grégory, F-06560 Valbonne, France*
[5] *Université Montpellier 2, Laboratoire Charles Coulomb, UMR 5221, F-34095 Montpellier, France*
[6] *CNRS, Laboratoire Charles Coulomb, UMR 5221, F-34095 Montpellier, France*

## Abstract

*In order to achieve polariton lasing at room temperature, a new fabrication methodology for planar microcavities is proposed: a ZnO-based microcavity in which the active region is epitaxially grown on an AlGaN/AlN/Si substrate and in which two dielectric mirrors are used. This approach allows as to simultaneously obtain a high-quality active layer together with a high photonic confinement as demonstrated through macro-, and micro-photoluminescence (µ-PL) and reflectivity experiments. A quality factor of 675 and a maximum PL emission at k=0 are evidenced thanks to µ-PL, revealing an efficient polaritonic relaxation even at low excitation power.*

[*]E-mail address: francois.reveret@lasmea.univ-bpclermont.fr

Zinc oxide microcavities have been widely studied during the few past years in order to obtain a laser device operating in the near ultraviolet or blue spectral range [1,2]. Thanks to the large binding energy (60 meV) and the small Bohr radius (1.8 nm) of ZnO excitons, the achievement of a polariton laser working at room temperature based on the coherent emission from exciton-polaritons in their ground state is expected [3,4]. A strong coupling regime (SCR) up to 410 K [2,5-9] is demonstrated but the lasing threshold is not crossed, mainly due to the low photon lifetime. Two different conceptions of ZnO microcavities have been reported: all oxide structures with high quality factors but polycrystalline zinc oxide layers [5-7], or hybrid cavities with a better quality of the active layer but a reduced photonic confinement [2,8-9]. Similar to previous works on GaN-based microcavities [10,11] a third approach can be followed using two dielectric mirrors after removal of the substrate used for the epitaxy of the active layer for a high quality factor and a reduced inhomogeneous broadening of the ZnO excitons.

In this letter, we present the first realization of a double dielectric mirror *epitaxial* ZnO-based microcavity working in the strong coupling regime. We first describe the growth process and the achievement of SCR at room temperature. The suppression of the bottleneck effect in the lower polariton branch at a low power density appears as proof of a long polaritonic lifetime.

The cavity structure is given in Fig. 1(a). First, a high-quality half-wavelength ($\lambda/2$) layer of ZnO was grown by molecular beam epitaxy (MBE) on a 3 AlN/(Al,Ga)N pair stack grown by MBE on a (111) Si substrate. On the top of the ZnO layer, a 3.5-pair $SiO_2/SiN_x$ dielectric Bragg reflector (DBR) was deposited by radio-frequency -plasma-enhanced chemical vapor deposition (RF-PECVD) and completed by a 200-nm-thick aluminum layer (theoretical reflectivity: 98.5%). The obtained structure was then bonded onto a Pyrex plate and the silicon substrate was subsequently removed by chemical etching terminated by $SF_6/O_2$ inductively coupled plasma etching. The first AlN layer is used as an etch-stop during the removal step of the silicon substrate. Finally, a 10-pair $SiO_2/SiN_x$ top Bragg mirror was deposited by RF-PECVD in place of the silicon substrate.

Angle resolved reflectivity (R) and photoluminescence (PL) experiments were performed at low (5 K) and room temperature with incidence angles from 2° to 85° (resolution < 1°). A halogen lamp is used for reflectivity. A 266 nm continuous wave (cw) laser is focused on a 300-µm-diameter spot (macro-PL measurements) and is adjustable from 2 to 20 µm for µ-PL though a UV objective (numerical aperture 0.4).

Prior to the deposition of the last dielectric mirror, the PL spectrum of the half cavity after Si etching was recorded at 10 K (Fig. 1(b)). An intense luminescence associated to neutral donor bound excitons in ZnO is observed at about 3.36 eV [12] whereas the peak at 3.90 eV corresponds to the excitonic luminescence of (Al,Ga)N [13]. Indeed, neither the quality of the ZnO layer nor that of the AlGaN of the nitride stack is affected by the removal of the substrate. Two photonic modes are also observed on the low energy side of the spectrum.

The quality factor of the complete cavity has been measured far from the ZnO excitonic energies in order to assess the linewidth of the uncoupled cavity mode. Using macro-PL excitation conditions, a quality factor (Q) of 150 is obtained, far from the theoretical prediction (Q=900) based on transfer matrix simulations, which consider solely a homogeneous broadening of the cavity mode. The Q-value measured by macro-PL is strongly reduced with respect to the theoretical one due to thickness fluctuations both in the active region and in the DBRs [14]. In our case, thickness fluctuations due to the inhomogeneous and incomplete removal of the Si substrate can further contribute to the inhomogeneous broadening. Indeed, a mapping of the cavity mode energy (Fig. 1(c)) shows that the spatial inhomogeneities occur at the scale of 2-10 µm. Thus, by reducing the µPL spot diameter to 2 μm we were able to differentiate extended areas where Q reaches 675, close to the theoretical value (Fig. 1(d)).

Angle-resolved reflectivity measurements recorded at room temperature in TM polarization (i.e., with the electric field parallel to the incidence plane) for an incidence angle from 5° to 70° are reported in Fig. 2(a). On the upper spectrum (5°), two features are observed: the optical cavity mode at 3200±1 meV negatively detuned and the broad excitonic transition at 3310±10 meV. With increasing incidence angle, the cavity mode moves towards the exciton energy and the SCR is evidenced by the anticrossing of the two modes. The resonance is observed at 35° and the Rabi splitting (Ω) is equal to 42±5 meV. Due to the current cavity design, the electric field is more intense in the nitride stack than in the ZnO active layer and consequently the Rabi splitting is smaller than could be expected for a conventional λ/2 thick ZnO layer. A second anticrossing is also observed between the LPB and the Bragg polariton (BP) between 45° and 50° (Ω=30±5 meV). The reflectivity measurements at 5 K (not shown here) also exhibit the SCR with a Rabi splitting equal to 55±4 meV in TM polarization. The increase of the Rabi splitting at low temperature can be simply explained by the reduction of the exciton linewidth, which approaches that of the photonic mode [15]. The detection at 5 K of A, B, and C excitons at 3376±3, 3382±3, and 3427±2 meV, respectively, in agreement with previous measurements [8], confirms the high quality of the active layer. The dispersion of the polaritonic modes deduced from reflectivity (solid squares) and PL (open circles) data at 300 K is plotted in Fig. 2(b). A very good agreement is found between both measurements. The quasiparticle (QP) model is used to calculate the polaritonic dispersion curves and the photonic and excitonic compositions of the modes [16]. Three quasiparticles are considered in the current cavity: the cavity (cav) and Bragg (b) modes, as well as one excitonic transition (exc), which accounts for A and B ZnO excitons, as they cannot be individually resolved at room temperature. Optical modes interact with the exciton through coupling terms called $V_{cav}$ and $V_b$. The damping terms of the oscillators associated with these quasiparticles and deduced from the analysis of the reflectivity spectra are $\Gamma_{exc}$=50 meV, $\Gamma_{cav}$=21 meV, and $\Gamma_b$=54 meV. In Fig. 2(b), the dotted lines represent the uncoupled excitonic transition and the dispersion of the photonic modes whereas the solid lines correspond to the calculations using the QP model. The coupling terms ($V_{cav}$ and $V_b$) deduced from the simulation, 22 and 14 meV, respectively, are consistent with the expected values (since $\Omega$~2V).

The macro-PL (illuminating spot of 300 µm in diameter) of the cavity has been recorded as a function of the emission angle ($\theta$) - with respect to the normal direction of the sample – for various excitation powers. The integrated intensity of the LPB measured for three laser intensities is divided by the corresponding value of the laser intensity (16, 40, and 170 W/cm²). Taking into account the photonic character of the LPB previously calculated (Fig. 2(c)), it is possible to determine the polariton distribution within the LPB, which is represented in Fig. 3(a). At low excitation intensity (16 W/cm²) the polariton distribution is maximum at about 18° due to the so-called bottleneck effect, which prevents thermalization of polaritons towards the k=0 low energy states. In this configuration, the relaxation time ($\tau_{rel}$) is probably larger than the polariton lifetime ($\tau_{pol}$) given that the photonic lifetime, which in the current situation is limiting the overall polariton lifetime, equals 0.03 ps due to the large spatial inhomogeneities [17]. Under an excitation power of 40 W/cm², this bottleneck can be overcome and the polariton distribution is approximatively constant whatever the emission angle. The suppression of the bottleneck can be attributed to the increase of the polariton-polariton scattering, which is more efficient when more polaritons are created [18, 19]. For the largest excitation power (170 W/cm²), polaritons not only relax more efficiently towards k=0 but also leak into the BP branch, as observed in ref. 18.

However, due to the spatial inhomogeneities of the Q-factor (i.e., of the photon lifetime), the macro-PL signal, and consequently the observed relaxation in Fig. 3(a), is averaged over regions of very different optical properties (regions of different Q-values and of different detunings, as illustrated by the blue spectrum in Fig. 1(d)). In order to get rid of this averaging, the luminescence of the cavity was also recorded as a function of the emission angle through Fourier imaging by using a microscope objective which allows us to collect PL emission up to 20°. The advantage of this method lies in the fact that it allows to investigate the angular dependence of the emission within small areas that are more homogeneous with respect to the cavity finesse. A region 20 µm in diameter was investigated and the results are reported in Figs. 3(b) and 3(c) for excitation intensities equal to 4.5 and 400 W/cm$^2$, respectively. The emission patterns are similar in both cases, with no bottleneck observable even at the lowest excitation intensity, a situation radically different to the one observed by macro-PL. With this spot diameter (20 µm), Q is found to be equal to 300 (photon lifetime=0.06 ps), i.e., twice as large as in the macro-PL measurements. It is noteworthy that even though the spectra in Fig. 3(b) and 3(c) were recorded at a position of larger negative detuning (-120 meV) than those in Fig. 3(a) and, therefore, with a larger $\tau_{rel}$, polaritons are thermalized at k=0, pointing clearly towards a polariton lifetime that is now equal to or larger than $\tau_{rel}$ [17]. Thus, when the investigated area is homogeneous in Q and with a Q-value larger than its mean value (Figs. 3(b) and 3(c)), efficient relaxation towards k=0 can be easily obtained (even at low excitation intensities), this being a necessary condition for polariton condensation. Unfortunately the polariton lifetime, mainly determined by its photonic lifetime, was still too short to enable polariton condensation.

In conclusion, we have fabricated an epitaxial ZnO microcavity embedded between two dielectric mirrors and studied its optical properties. This approach exploits the etching selectivity between silicon and nitrides, and allows us to increase the quality factor of the cavity while maintaining a high optical quality of the active layer. However, as evidenced by μ-PL experiments at room temperature, the cavity Q is spatially inhomogeneous on a scale of several micrometers. Larger excitation power densities are therefore required in order to overcome the bottleneck effect when exciting larger regions, as compared to the spatial scale of the Q inhomogeneities. Finally, efficient polariton relaxation towards k=0 could be accomplished at room temperature in regions several micrometers in diameter at low excitation intensities.

The authors acknowledge G. Malpuech for fruitful discussions on the polaritonic behavior. This work has been supported by the ANR under the "ZOOM" project (Grant No. ANR-06-BLAN-0135) and the European Union under the "Clermont4" project (Grant No. 235114).

**Figure captions**

Fig. 1. (a) Sketch of the full hybrid ZnO microcavity (after processing). (b) Photoluminescence (T=10 K, 244 nm cw excitation) of the structure after Si removal and before the deposition of the top DBR. (c) Energy mapping in a small area (30 µm) using a cw 266 nm laser (d) Micro-photoluminescence (spot of 2 µm in diameter) at 300 K recorded on a region of homogeneous Q and cavity mode energy (red spectrum) and at the boundary between two regions of different Q and cavity mode energy values (blue spectrum).

Fig. 2. (a) Evolution of angle-resolved reflectivity from 5° to 70° recorded in TM polarization at room temperature. (b) Dispersion of the polaritonic modes deduced from reflectivity (solid squares) and photoluminescence (open circles) data. The solid lines correspond to the calculation using the QP model; dotted lines correspond to the uncoupled modes and the excitonic dispersion. (c) Angular dependence of the eigenstate composition of the LPB calculated within the QP model. Note that the excitonic part is rather small below 20° (negative detuning) and beyond 50° (positive detuning, due to the arrival of the Bragg mode).

Fig. 3. (a) Distribution of the polaritons in the LPB is deduced from angle-resolved macro-photoluminescence (spot of 300 µm in diameter) at 300 K for different excitation powers (16, 40, and 170 W/cm²). (b,c) Images of the angular dispersion of the LPB with a cw 266 nm laser focused on a 20 µm diameter spot for two excitation intensities of 4.5 and 400 W/cm².

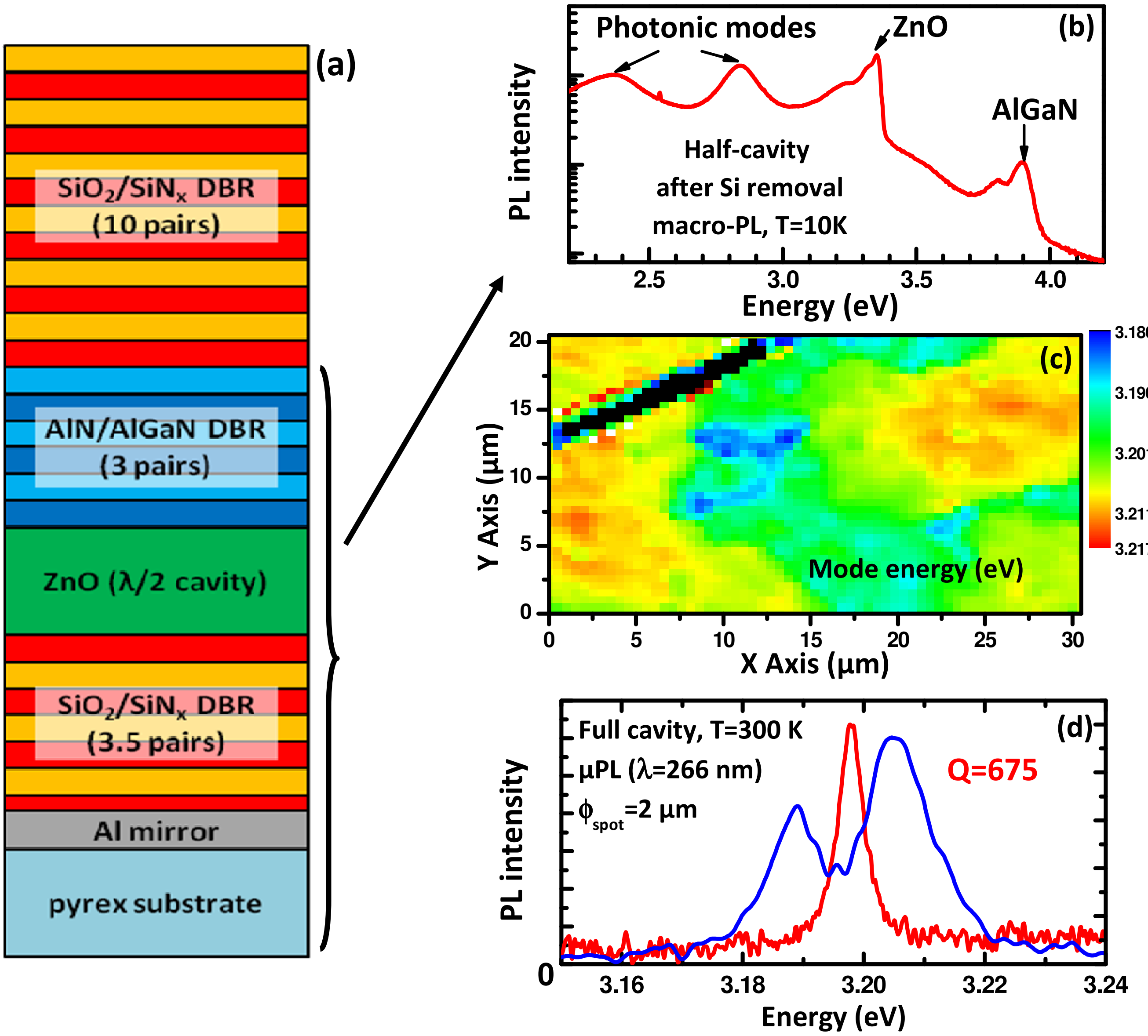

Figure 1

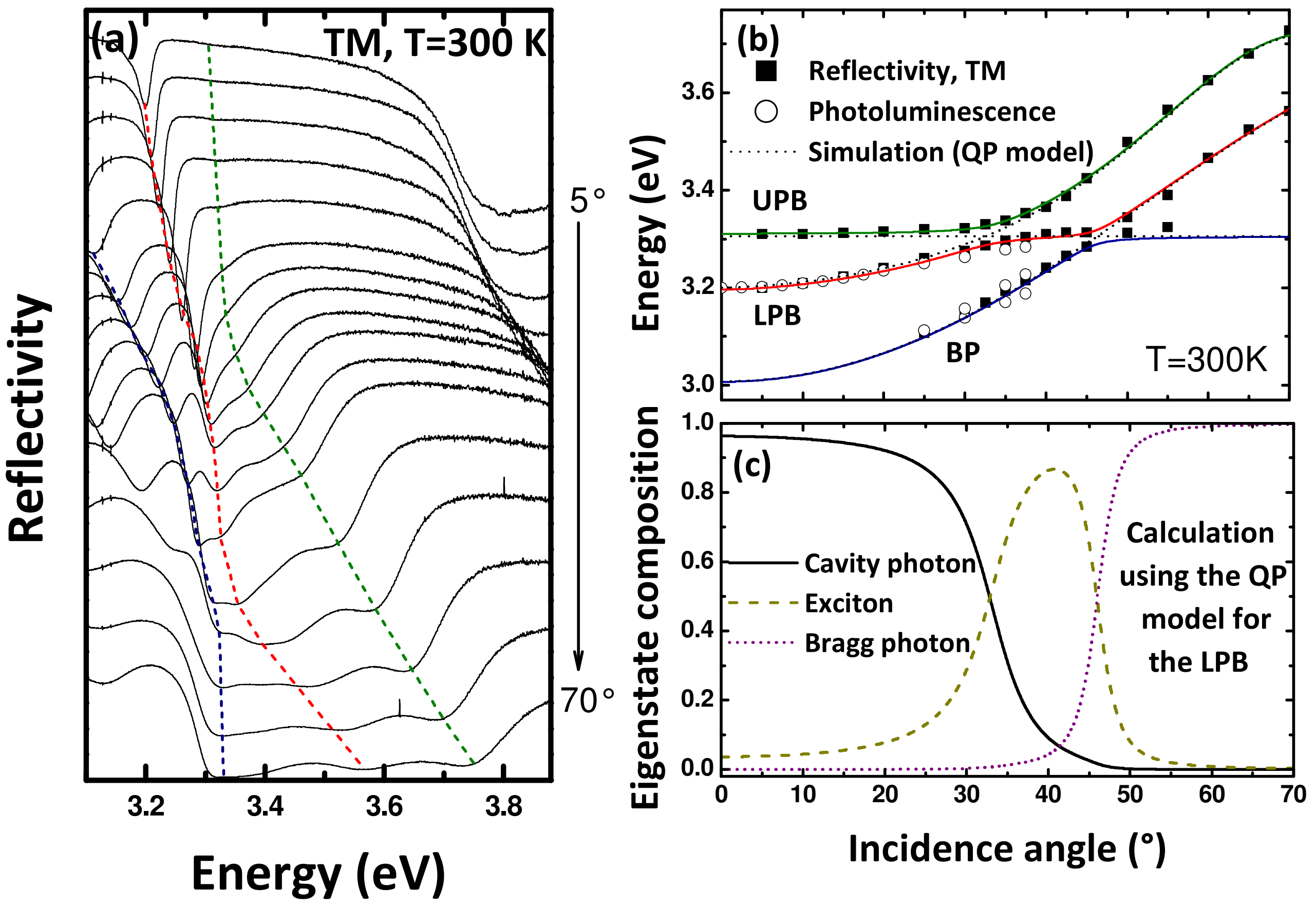

(a)
TM, T=300 K
Reflectivity
5°
70°
3.2
3.4
3.6
3.8
Energy (eV)
(b)
Reflectivity, TM
Photoluminescence
Simulation (QP model)
UPB
LPB
BP
T=300K
Energy (eV)
3.0
3.2
3.4
3.6
(c)
Cavity photon
Exciton
Bragg photon
Calculation using the QP model for the LPB
Eigenstate composition
0.0
0.2
0.4
0.6
0.8
1.0
0
10
20
30
40
50
60
70
Incidence angle (°)

Figure 2

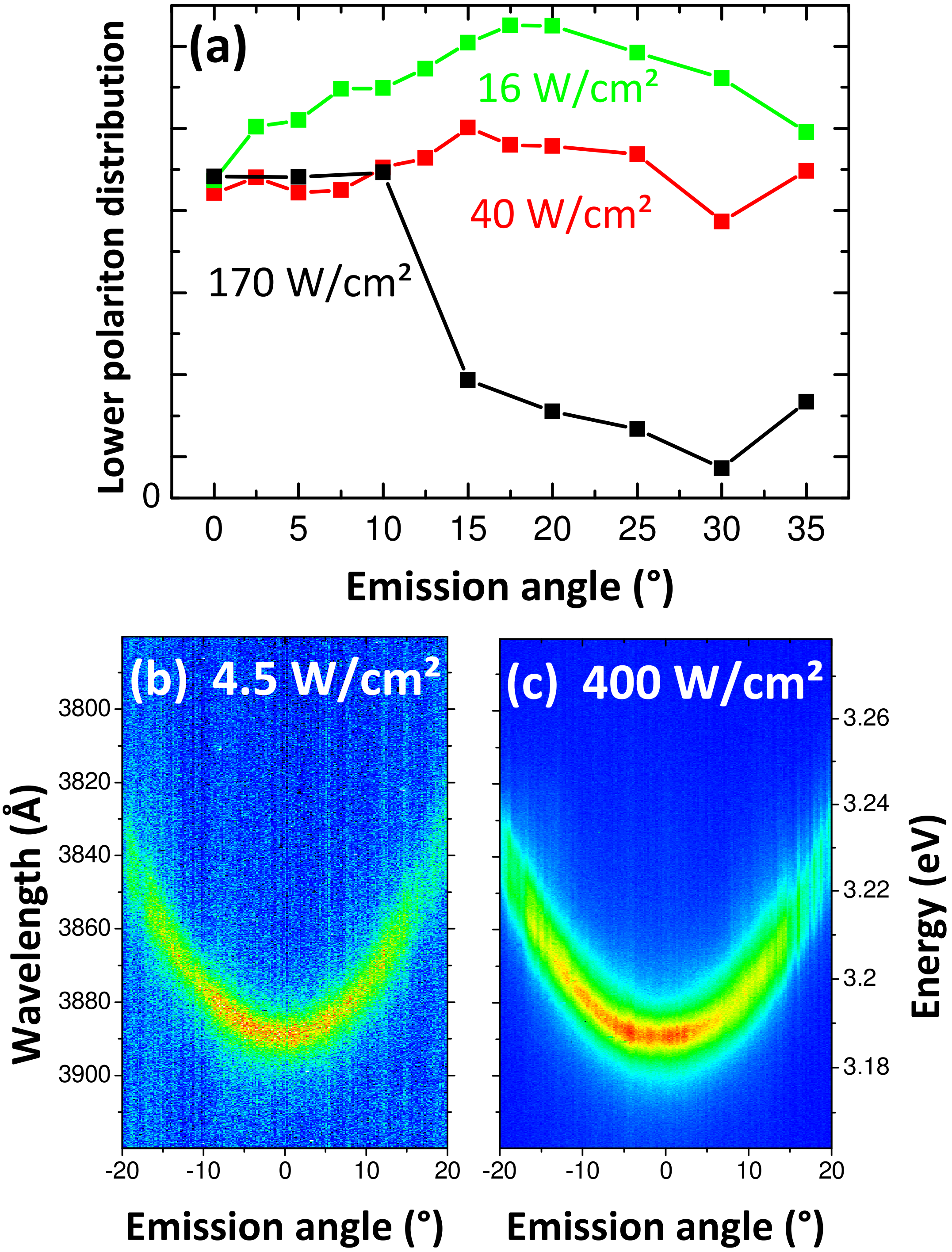

Figure 3